\documentclass[journal]{IEEEtran}
\ifCLASSINFOpdf

\else

\fi
\usepackage{mathrsfs}

\usepackage{amsmath}
\usepackage{ntheorem}

\usepackage{makeidx}  
\usepackage{algorithm}
\usepackage{algorithmic}
\usepackage{graphicx}
\usepackage{subfigure}
\usepackage{epstopdf}
\usepackage{bm}
\usepackage{cite}
\usepackage{stfloats}

\usepackage{amssymb}
\usepackage{graphicx}

\usepackage{url}

\usepackage{tabularx,booktabs}
\newcolumntype{C}{>{\centering\arraybackslash}X} 
\usepackage{lipsum}

\usepackage{makecell} 

\usepackage{graphicx}
\usepackage{color}

\newtheorem{lem}{Lemma}

\newtheorem{corollary}{Corollary}

\definecolor{dblue}{RGB}{15,89,164}

\begin{document}
\title { Ergodic Rate Analysis of Reconfigurable Intelligent Surface-Aided Massive MIMO Systems with ZF Detectors }
\author{Kangda Zhi, Cunhua Pan, Hong Ren and Kezhi Wang\thanks{ \itshape (Corresponding author: Cunhua Pan.)\upshape
		 		
		K. Zhi, C. Pan are with the School of Electronic Engineering and Computer Science at Queen Mary University of London, London E1 4NS, U.K. (e-mail: k.zhi, c.pan@qmul.ac.uk).
		
		H. Ren is with the National Mobile Communications Research Laboratory, Southeast University, Nanjing 210096, China. (hren@seu.edu.cn).

K. Wang is with Department of Computer and Information Sciences, Northumbria University, UK. (e-mail: kezhi.wang@northumbria.ac.uk).}
\vspace{-20pt}
}

\maketitle

\begin{abstract}
	This letter investigates the reconfigurable intelligent surface (RIS)-aided massive multiple-input multiple-output (MIMO) systems with a two-timescale design.
	First, the zero-forcing (ZF) detector is applied at the base station (BS) based on instantaneous aggregated channel state information (CSI), which is the superposition of the direct channel and the cascaded user-RIS-BS channel.
	Then, by leveraging the channel statistical property, we derive the closed-form ergodic achievable rate expression. Using a gradient ascent method, we design the RIS passive beamforming relying only on the long-term statistical CSI. We prove that the ergodic rate {scales} on the order of $\mathcal{O}\left(\log_{2}\left(MN\right)\right)$, where $M$ and $N$ denote the number of BS antennas and RIS elements, respectively. We also prove the striking superiority of the considered RIS-aided system with ZF detectors over the RIS-free systems and RIS-aided systems with maximum-ratio combining (MRC).
	
\end{abstract}


\IEEEpeerreviewmaketitle

\section{Introduction}
As an emerging technique, reconfigurable intelligent surface (RIS) has been widely investigated and recognized as a cost-effective complement for future systems\cite{pan2020intelligent,pan2020multicell}.
 The RIS mainly consists of a large number of passive reflecting elements that have low hardware cost and energy consumption. Besides, RIS can help conventional systems overcome the blockage issue and assist the transmission by creating high-quality transmission paths.

A well-acknowledged challenge for the RIS is that it may introduce heavy channel estimation overhead. Fortunately, a novel and more practical countermeasure, named as two-timescale beamforming design, has been proposed and validated by some contributions{\cite{han2019large,kammoun2020asymptotic,zhao2021twoTimeScale,jia2020analysis,gao2021LMMSE,papa2021imperfect,papa2021maxmin}}. On one hand, the two-timescale scheme aims at designing the passive RIS beamforming based on purely statistical channel state information (CSI), and this could effectively reduce the overhead and the energy consumption in the operation of the RIS\cite{zhao2021twoTimeScale}. On the other hand, the two-timescale scheme designs the BS beamforming based on the instantaneous aggregated channel, which is the superposition of the direct channel and cascaded user-RIS-BS channel. As a result, the aggregated channel has the same dimension as that in conventional systems. Thus, two-timescale schemes possess the same channel estimation overhead as conventional systems.

Inspired by the above benefits, the two-timescale design has been recently exploited in RIS-aided massive MIMO systems\cite{zhi2020directLinks}. It has demonstrated that by integrating an RIS into conventional massive MIMO systems, the rate performance can be significantly improved, especially when the original direct links are weak due to the blockage. However, only the simple maximum-ratio combining (MRC) detector was considered in \cite{zhi2020directLinks}, and it was revealed that the achieved gains are limited by the multi-user interference. Therefore, it is expected that the zero-forcing (ZF) detector, which can effectively mitigate  this interference, is more suitable for RIS-aided massive MIMO systems. Different from \cite{zhi2020directLinks}, when using ZF detectors, matrix inversion operator introduces the additional technical challenges of deriving the ergodic capacity.

Against the above background, in this letter, we consider an RIS-aided massive MIMO system with ZF detectors. We derive the closed-form expression for the ergodic rate, which only relies on the long-term CSI. We then design the RIS based on a gradient ascent algorithm. By analyzing the rate expression, we find that it scales on the order of $\mathcal{O}\left(\log_{2}\left(MN\right)\right)$, which indicates that the RIS-aided massive MIMO system with ZF detectors has the ability to achieve ultra-high system capacity.

\section{System Model}
\begin{figure}
	\setlength{\abovecaptionskip}{0pt}
	\setlength{\belowcaptionskip}{-20pt}
	\centering
	\includegraphics[width= 0.35\textwidth]{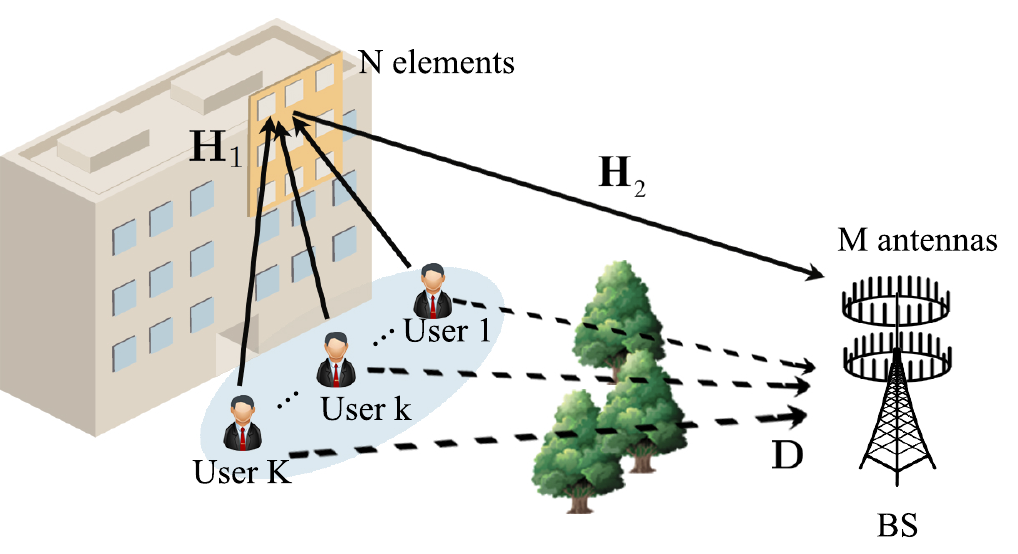}
	\DeclareGraphicsExtensions.
	\caption{Uplink transmission in RIS-aided massive MIMO systems.}
	\label{figure1}
	\vspace{-10pt}
\end{figure}

As illustrated in Fig. \ref{figure1}, the uplink transmission of a massive MIMO system is considered.
 Different from conventional systems,
 an RIS is introduced and equipped at the facade of a tall building close to $K$ single-antenna users to improve their channel conditions.
 The considered model is especially suitable for the scenario where some cell-edge users suffer from service degradation. Denote the number of BS antennas and RIS elements as $M$ and $N$, respectively, where $M>K$. Then, we can define the channel between the users and the RIS, the channel between the RIS and the BS, the direct channel between the users and the BS as $\mathbf{H}_1 \in \mathbb{C}^{N\times K}$, $\mathbf{H}_2 \in \mathbb{C}^{M\times N}$ and $\mathbf{D} \in \mathbb{C}^{M\times K}$, respectively.

Define the phase shift matrix of the RIS as $\mathbf{\Phi} = {\rm diag}\left\{      e^{j \theta_1},\ldots,e^{j \theta_N}             \right\}$ where $\theta_n$ is the phase shift of the $n$-th RIS element. Herein, we can express the cascaded user-RIS-BS channel as $\mathbf{G} = \mathbf{H}_2\mathbf{\Phi}\mathbf{H}_1 \in \mathbb{C}^{M\times K}$, and then express the aggregated channel from users to the BS as $\mathbf{Q} = \mathbf{G}+\mathbf{D}\in \mathbb{C}^{M\times K} $. It is worth noting that this aggregated channel $\mathbf{Q}$ possesses the same dimension as conventional massive MIMO systems.

Based on the above definitions, we next present the detailed channel model for $\mathbf{Q}$. Firstly, considering that the direct links may be easily blocked\cite{han2019large}, we adopt the Rayleigh channel model for $\mathbf{D}$ as follows
\begin{align}\label{D}
	\mathbf{D}=\tilde{\mathbf{D}} \boldsymbol{\Omega}_d^{1 / 2},
\end{align}
where $ \boldsymbol{\Omega}_d=\operatorname{diag}\left\{\gamma_{1}, \ldots, \gamma_{K}\right\} $, and $\gamma_k$ denotes the distance-dependent path-loss factor. Each element of matrix $ \tilde{\mathbf{D}} \in \mathbb{C}^{M\times K}$ is independent and identically distributed (i.i.d.) complex Guassian random variables, whose mean is zero and variance is unit.

Next, since we consider that the RIS is deployed close to the users, and according to the fact that the RIS is often installed above the ground, we assume that the user-RIS channels have purely line-of-sight (LoS) paths. Then, we denote
\begin{align}\label{H1}
\mathbf{H}_{1}=\left[\sqrt{\alpha_{1}} \,\overline{\mathbf{h}}_{1}, \ldots, \sqrt{\alpha_{K}}\, \overline{\mathbf{h}}_{K}\right],
\end{align}
where ${\alpha_k}$ denotes the path loss. To specify the LoS channel $\overline{\mathbf{h}}_{k}$, we utilize the two-dimensional uniform squared planar array (USPA) model\cite{zhou2020hardware}. Then, the array response vector for a $\sqrt{X}\times \sqrt{X}$ USPA can be expressed as follows
\begin{align}\label{USPA}
&{{\bf a}_X}\left( {\vartheta _{}^a,\vartheta _{}^e} \right) = \left[ 1,...,{e^{j2\pi \frac{d}{\lambda }\left( {x\sin \vartheta _{}^e\sin \vartheta _{}^a + y\cos \vartheta _{}^e} \right)}},\right.\nonumber\\
&\qquad\;\;\left. \ldots,{e^{j2\pi \frac{d}{\lambda }\left( {\left( {\sqrt X  - 1} \right)\sin \vartheta _{}^e\sin \vartheta _{}^a + \left( {\sqrt X  - 1} \right)\cos \vartheta _{}^e} \right)}} \right]^T,
\end{align}
where $0\leq x,y\leq \sqrt{X}-1$ are element indices in the two-dimensional planar array, $d$ and $\lambda$ denote the element spacing and wavelength, $\vartheta _{}^a$ and $\vartheta _{}^e$ are azimuth and elevation angles in the propagation path, respectively. Therefore, denoting by the azimuth and elevation AoA of user $k$ as $\varphi_{k r}^{a}$ and $\varphi_{k r}^{e}$, we can now express $\overline{\mathbf{h}}_{k} = {{\bf a}_N}\left( {\varphi_{k r}^{a}, \varphi_{k r}^{e}} \right) $.

Since the RIS is placed near the users, the distance between the RIS and the BS could be a bit large. Even though both the RIS and the BS have certain heights, it is still not guaranteed that the RIS-BS channel is purely LoS. As a result, the Rician model is suitable for the considered RIS-BS channel. Besides, by adjusting the value of Rician factors, we can study the impacts of scatterers in RIS-aided systems. This feature is important, since many works have proven that rich scattering environment is beneficial in conventional massive MIMO systems\cite{jin2007rank1}, while the corresponding impact in RIS-aided massive MIMO systems with ZF detector is still unknown. Thus, we define
\begin{align}
\begin{array}{l}
\mathbf{H}_{2}=\sqrt{\frac{\beta \delta}{\delta+1}} \overline{\mathbf{H}}_{2}+\sqrt{\frac{\beta}{\delta+1}} \tilde{\mathbf{H}}_{2},
\end{array}
\end{align}
where $\beta$ is the path loss, $\delta$ represents the Rician factor. Note that the Rician factor, varying from $0$ to $\infty$, characterizes the strength ratio between LoS and non-LoS (NLoS) paths. The NLoS path $\tilde{\mathbf{H}}_2$ contains i.i.d. complex Gaussian random variables with zero mean and unit variance. Recalling USPA model (\ref{USPA}), the LoS path, $\overline{\mathbf{H}}_2$, is written as
\begin{align}\label{H2}
\overline{\mathbf{H}}_2= \mathbf{a}_{M}\left(\phi_{r}^{a}, \phi_{r}^{e}\right) \mathbf{a}_{N}^{H}\left(\varphi_{t}^{a}, \varphi_{t}^{e}\right) \triangleq \mathbf{a}_{M}\mathbf{a}_{N}^H,
\end{align}
where a notational simplification $\overline{\mathbf{H}}_2\triangleq \mathbf{a}_{M}\mathbf{a}_{N}^H $ is applied in the sequel of this paper. Note that the rank of matrix $\overline{\mathbf{H}}_2$ is one. This means that when $\delta\to\infty$, the cascaded channel $\mathbf{G}$ may become rank-deficient, and then the achievable spatial multiplexing gains may degrade.

We can now express the $M\times 1$ received signal vector at the BS as
\begin{align}
\mathbf{y}=\sqrt{p} \mathbf{Q} \mathbf{x}+\mathbf{n}=\sqrt{p} \left(\mathbf{H}_2\mathbf{\Phi}\mathbf{H}_1 +\mathbf{D}\right) \mathbf{x}+\mathbf{n},
\end{align}
where $\mathbf{x}=\left[x_1,\ldots,x_K\right]^T\sim \mathcal{CN}\left(\mathbf{0},\mathbf{I}_K\right)$ includes the transmit symbols from $K$ users, and $\mathbf{n}\sim \mathcal{CN}\left(\mathbf{0},\sigma^2\mathbf{I}_M\right)$ is the noise vector. For simplicity, we assume that all users transmit with the same power $p$.

	To facilitate the analysis, we assume that in each channel coherence time, the instantaneous aggregated channel $\mathbf{Q}$ is perfectly known at the BS, which serves as an upper bound for practical systems. Based on the two-timescale design framework, we need to design the BS beamforming based on instantaneous aggregated CSI, i.e., $\mathbf{Q}$. Thus, the ZF detector at the BS is designed as
$ 	\mathbf{A}=\mathbf{Q}\left(\mathbf{Q}^{H} \mathbf{Q}\right)^{-1}$,
which results in $\mathbf{A}^{H} \mathbf{Q} = \mathbf{I}_K$.
Thus, the detected symbol vector is given by
\begin{align}\label{rr}
\begin{aligned}
\mathbf{r}&=\mathbf{A}^{H} \mathbf{y} =\sqrt{p} \mathbf{x}+\left(\mathbf{Q}^{H} \mathbf{Q}\right)^{-1} \mathbf{Q}^{H} \mathbf{n}.
\end{aligned}
\end{align}

For ZF, there is no multi-user interference. As a result, the signal-to-interference-plus-noise ratio (SINR) reduces to the ratio of transmit power and noise. Based on (\ref{rr}), the SINR of user $k$ is given by
	\begin{align}
\mathrm{SINR}_k &=\frac{p\left[\mathbb{E}_{\mathbf{x}}\left\{\mathbf{x}\mathbf{x}^H\right\}\right]_{kk}}
{[\mathbb{E}_{\mathbf{n}}\{
	(\mathbf{Q}^{H} \mathbf{Q})^{-1} \mathbf{Q}^{H} \mathbf{n}
	\mathbf{n}^H\mathbf{Q} (\mathbf{Q}^{H} \mathbf{Q})^{-1}
	\}]_{kk}}\nonumber\\
&=\frac{p}{\sigma^{2}\left[(\mathbf{Q}^{H} \mathbf{Q})^{-1}\right]_{k k}}.
	\end{align}

Then, the $k$-th user's ergodic rate is lower bounded by
\begin{align}\label{rate_definitation}
R_{k}&=\mathbb{E}  \left\{ \log _{2} \left(  1+  \mathrm{SINR}_k\right)\right\} \\\label{lb}
&\overset{(a)}{\geq} \log _{2}\left(1+\frac{p}{\sigma^{2} \mathbb{E}\left\{\left[(\mathbf{Q}^{H} \mathbf{Q})^{-1}\right]_{k k}\right\}}\right),
\end{align}
where $(a)$ utilizes the Jensen's inequality based on the fact that  function $f(x)\!=\!\log_{2}\left(1+\frac{1}{x}\right)$ is convex with respect to $x$.


\section{Rate Analysis and RIS Design}
In this section, we first derive the closed-form expression for the rate $R_k$, and then use the derived expression to propose a statistical CSI-based RIS design.

To derive $R_k$, we need to compute $  \mathbb{E}\left\{\left[(\mathbf{Q}^{H} \mathbf{Q})^{-1}\right]_{k k} \right\}   $. To this end, we expand matrix $\mathbf{Q}^H$ as
\begin{align}
	\begin{array}{l}
	\mathbf{Q}^{H}=\sqrt{\frac{\beta \delta}{\delta+1}} \mathbf{H}_{1}^{H} \mathbf{\Phi}^{H} \overline{\mathbf{H}}_{2}^{H}+\sqrt{\frac{\beta}{\delta+1}} \mathbf{H}_{1}^{H} \mathbf{\Phi}^{H} \tilde{\mathbf{H}}_{2}^{H}+\mathbf{\Omega}_{d}^{1 / 2} \tilde{\mathbf{D}}^{H}.
	\end{array}
\end{align}

For ease of exposition, we define $ \mathbf{Q}^{H} \triangleq\left[\mathbf{q}_{1}, \ldots, \mathbf{q}_{M}\right] $, $ \overline{\mathbf{H}}_{2}^{H} \triangleq\left[\overline{\mathbf{c}}_{1}, \ldots, \overline{\mathbf{c}}_{M}\right] $, $ \tilde{\mathbf{H}}_{2}^{H} \triangleq\left[\tilde{\mathbf{c}}_{1}, \ldots, \tilde{\mathbf{c}}_{M}\right] $, and $ \tilde{\mathbf{D}}^{H} \triangleq\left[\tilde{\mathbf{d}}_{1}, \ldots, \tilde{\mathbf{d}}_{M}\right] $. Recalling that $\tilde{\mathbf{H}}_2$ and $\tilde{\mathbf{D}}$ are all comprised of i.i.d. complex Gaussian variables, and $\tilde{\mathbf{H}}_2$ and $\tilde{\mathbf{D}}$ are mutual independent, we therefore have
\begin{align}
	&\tilde{\mathbf{c}}_{m} \sim \mathcal{C N}\left(\mathbf{0}, \mathbf{I}_{N}\right), 1 \leq m \leq M, \\
	&\tilde{\mathbf{d}}_{m} \sim \mathcal{C N}\left(\mathbf{0}, \mathbf{I}_{K}\right), 1 \leq m \leq M,
\end{align}
where $\tilde{\mathbf{c}}_{i}$ and $\tilde{\mathbf{c}}_{j}$ are mutual independent for $i\neq j$; $\tilde{\mathbf{d}}_{i}$ and $\tilde{\mathbf{d}}_{j}$ are mutual independent, for $i\neq j $; $\tilde{\mathbf{c}}_{i}$ and $\tilde{\mathbf{d}}_{j}$ are mutual independent for all $i$ and $j$. Then, since the linear transformation for a standard Gaussian random vector is still a Gaussian random vector\cite{muirhead2009aspects}, we obtain
\begin{align}
\begin{array}{l}
\sqrt{\frac{\beta \delta}{\delta+1}} \mathbf{H}_{1}^{H} \mathbf{\Phi}^{H} \overline{\mathbf{c}}_{m}+\sqrt{\frac{\beta}{\delta+1}} \mathbf{H}_{1}^{H} \mathbf{\Phi}^{H} \tilde{\mathbf{c}}_{m} \\
\sim \mathcal{C N}\left(\sqrt{\frac{\beta \delta}{\delta+1}} \mathbf{H}_{1}^{H} \mathbf{\Phi}^{H} \overline{\mathbf{c}}_{m}, \frac{\beta}{\delta+1} \mathbf{H}_{1}^{H} \mathbf{H}_{1}\right) ,\forall m
\end{array}
\end{align}
and $\mathbf{\Omega}_{d}^{1 / 2} \tilde{\mathbf{d}}_{m} \sim \mathcal{C N}\left(\mathbf{0}, \mathbf{\Omega}_{d}\right),\forall m$, where the facts $\mathbf{\Phi}^H\mathbf{\Phi}=\mathbf{I}_N$ and $\mathbf{\Omega}_d=\mathbf{\Omega}_d^H$ were used.

Next,
 taking into account that the sum of independent Gaussian vectors is still Gaussian distributed\cite[Theorem 1.2.14]{muirhead2009aspects}, we can obtain the statistics of the $m$-th column of aggregated channel $\mathbf{Q}^H$ as follows
\begin{align}
\begin{array}{l}
\mathbf{q}_{m}  \sim  \mathcal{C} \mathcal{N}\left(\sqrt{\frac{\beta \delta}{\delta+1}} \mathbf{H}_{1}^{H} \mathbf{\Phi}^{H} \overline{\mathbf{c}}_{m}, \;\frac{\beta}{\delta+1} \mathbf{H}_{1}^{H} \mathbf{H}_{1}+\boldsymbol{\Omega}_{d}\right),
\end{array}
\end{align}
where $\mathbf{q}_m$, $1\leq m \leq M$, are mutual independent. Therefore, $\mathrm{vec}\left(\mathbf{Q}^H\right)$ is a complex Gaussian vector with the following mean and covariance matrices
\begin{align}
\begin{array}{l}
\mathbb{E}\left\{{\mathrm{vec}}\left(\mathbf{Q}^{H}\right)\right\}={\mathrm{vec}}\left(\sqrt{\frac{\beta \delta}{\delta+1}} \mathbf{H}_{1}^{H} \mathbf{\Phi}^{H} \overline{\mathbf{H}}_{2}^{H}\right), \\
\mathrm{Cov}\left\{{\mathrm{vec}}\left(\mathbf{Q}^{H}\right)\right\}=\mathbf{I}_{M} \otimes\left(\frac{\beta}{\delta+1} \mathbf{H}_{1}^{H} \mathbf{H}_{1}+\mathbf{\Omega}_{d}\right),
\end{array}
\end{align}
where $\mathrm{vec}$ and $\otimes$ denote the vectorization by column stacking and Kronecker product, respectively.

Then, using the distribution of $\mathrm{vec}\left(\mathbf{Q}^H\right)$ and following the notations in \cite[Page 2]{jin2007rank1}, matrix $\mathbf{Q}$ is a complex Gaussian distributed matrix, written as
\begin{align}
\begin{array}{l}
\mathbf{Q} \sim \mathcal{C} \mathcal{N}\!\left(\!\sqrt{\frac{\beta \delta}{\delta+1}} \overline{\mathbf{H}}_{2} \mathbf{\Phi} \mathbf{H}_{1}, \mathbf{I}_{M}\! \otimes\!\left(\frac{\beta}{\delta+1} \mathbf{H}_{1}^{H} \mathbf{H}_{1}\!+\!\mathbf{\Omega}_{d}\!\right)\right)\!.\!\!\!
\end{array}
\end{align}

Therefore, the product $\mathbf{Q}^{H} \mathbf{Q}$ has a complex non-central  Wishart distribution\cite[Definition 10.3.1]{muirhead2009aspects}, which can be repressed as
\begin{align}\label{non_central_W}
\begin{array}{l}
\mathbf{Q}^{H} \mathbf{Q} \sim \mathcal{W}_{K}\left(M, \frac{\beta}{\delta+1} \mathbf{H}_{1}^{H} \mathbf{H}_{1}+\mathbf{\Omega}_{d} \right. \\
\;\;\quad,\left. (\frac{\beta}{\delta+1} \mathbf{H}_{1}^{H} \mathbf{H}_{1}+\mathbf{\Omega}_{d})^{-1}\! \frac{\beta \delta}{\delta+1} \mathbf{H}_{1}^{H} \mathbf{\Phi}^{H} \overline{\mathbf{H}}_{2}^{H} \overline{\mathbf{H}}_{2} \mathbf{\Phi} \mathbf{H}_{1}\!\right)\!.\!\!\!
\end{array}
\end{align}

Even though the non-central Wishart distribution (\ref{non_central_W}) is accurate, its statistics are very complicated, and then we cannot obtain a tractable expression for insightful analysis. To facilitate the analysis, as in contributions \cite{Cons2012ZF,Cons2015schur,zhang2014power}, we next approximate the non-central Wishart distribution (\ref{non_central_W}) as a central Wishart distribution with the same first-order moment.

To begin with, the first-order moment for the considered non-central Wishart distribution is\cite[Eq. (45)]{Cons2015schur}
\begin{align}
\begin{aligned}
&\mathbb{E}\left\{\mathbf{Q}^{H} \mathbf{Q}\right\}\\
&=M\left(\frac{\beta}{\delta+1} \mathbf{H}_{1}^{H} \mathbf{H}_{1}+\mathbf{\Omega}_{d}\right)+\frac{\beta \delta}{\delta+1} \mathbf{H}_{1}^{H} \mathbf{\Phi}^{H} \overline{\mathbf{H}}_{2}^{H} \overline{\mathbf{H}}_{2} \mathbf{\Phi} \mathbf{H}_{1} \\
&=M\left(\frac{\beta}{\delta+1} \mathbf{H}_{1}^{H} \mathbf{H}_{1}+\mathbf{\Omega}_{d}\right)+M \frac{\beta \delta}{\delta+1} \mathbf{H}_{1}^{H} \mathbf{\Phi}^{H} \mathbf{a}_{N} \mathbf{a}_{N}^{H} \mathbf{\Phi} \mathbf{H}_{1},
\end{aligned}
\end{align}
where the last equality is obtained by using ($\ref{H2}$) and $\mathbf{a}_M^{H}\mathbf{a}_M=M$.

Therefore, a virtual central Wishart distribution with this moment is given by\cite[Sec. $ \rm\uppercase\expandafter{\romannumeral5} $. A]{Cons2015schur}
\begin{align}\label{central_Wishart_QQ}
\begin{array}{l}
\mathbf{Q}^{H} \mathbf{Q} \!\sim\! \mathcal{W}_{K}\!\!\left(\!M, \frac{\beta}{\delta+1} \mathbf{H}_{1}^{H} \mathbf{H}_{1}\!+\!\mathbf{\Omega}_{d}\!+\!\frac{\beta \delta}{\delta+1} \mathbf{H}_{1}^{H} \mathbf{\Phi}^{H} \mathbf{a}_{N} \mathbf{a}_{N}^{H} \mathbf{\Phi} \mathbf{H}_{1}\!\right)\!.
\end{array}
\end{align}

Based on the obtained complex central Wishart distribution (\ref{central_Wishart_QQ}), with the help of \cite[Table $ \rm\uppercase\expandafter{\romannumeral1} $]{tague1994expectations}, we can obtain the expectation of the matrix inverse as follows
\begin{align}\label{inverse_exp}
\mathbb{E}\!\!\left\{\!\!\left(\mathbf{Q}^{H} \mathbf{Q}\right)^{-1}\!\right\}\!\!=\!\frac{\!\!\left(\!\frac{\beta}{\delta+1} \mathbf{H}_{1}^{H} \mathbf{H}_{1}\!+\!\mathbf{\Omega}_{d}\!+\!\frac{\beta \delta}{\delta+1} \mathbf{H}_{1}^{H} \mathbf{\Phi}^{H} \mathbf{a}_{N} \mathbf{a}_{N}^{H} \mathbf{\Phi} \mathbf{H}_{1}\!\!\right)^{-1}\!}{M-K}\!.
\end{align}

Substituting (\ref{inverse_exp}) into (\ref{lb}), we obtain the lower bound of the ergodic rate of user $k$ as follows
\begin{align} \label{rate}
R_{k} \!\geq\!  \log _{2} \!\!\left(\!1 \!+ \! \frac{p\left(M-K\right) }{\sigma^{2}(\delta\! + \!1) \! \left[(\mathbf{\Lambda}+ \beta\delta \mathbf{H}_{1}^{H} \mathbf{\Phi}^{H} \mathbf{a}_{N} \mathbf{a}_{N}^{H} \mathbf{\Phi} \mathbf{H}_{1})^{-1}\right]_{kk}}   \right)\!,
\end{align}
where $\mathbf{\Lambda}=\beta \mathbf{H}_{1}^{H} \mathbf{H}_{1}+(\delta+1) \mathbf{\Omega}_{d}$.

Note that the derived expression, (\ref{rate}), depends only on the statistical CSI, since the instantaneous CSI-related variables have been averaged out.
Therefore, based on the two-timescale design framework, we can use (\ref{rate}) to design the phase shifts of the RIS only relying on statistical CSI.
Since the statistical CSI-based phase shifts design only needs to be done on a large time-scale, the overhead can be effectively reduced. Besides, it is clear that (\ref{rate}) is an increasing function of $p$ and the RIS-BS channel strength $\beta$, but it is a decreasing function of noise power $\sigma^2$.
\begin{corollary}\label{corollary1}
	As $M\to\infty $, the rate can maintain non-zero when the power is scaled down proportionally to $p=1/M$.
\end{corollary}

\itshape \textbf{Proof:}  \upshape It can be proved by noticing that all the matrices in the denominator of (\ref{rate}) do not depend on $M$.  \hfill $\blacksquare$

\begin{corollary}\label{corollary2}
	When $M\to\infty$ or $p\to\infty$, RIS-aided massive MIMO systems with ZF detectors perform much better than that with MRC detectors.
\end{corollary}

 \itshape \textbf{Proof:}  \upshape Based on (\ref{rate}), when $M\to\infty$ or $p\to\infty$, we have $R_k\to\infty$, while the rate in RIS-aided massive MIMO systems with MRC detectors is still bounded due to the multi-user interference, as proved in \cite[Eq. (7)]{zhi2020directLinks}. \hfill $\blacksquare$

\begin{corollary}\label{corollary3}
	When $\beta=0$, i.e., without the existence of the RIS, the rate of user $k$ reduces to
	\begin{align}\label{rate_conventional}
	R_{k} \geq
	\log _{2}\left(1+{p{\left(M-K\right)\gamma_k}}/{\sigma^{2}} \right),
	\end{align}
	which is the same rate as \cite[Eq. (20)]{ngo2013energy}, and scales on the order of $\mathcal{O}\left(\log_{2}\left(M\right)\right)$.
\end{corollary}

\begin{corollary}\label{corollary4}
The ergodic rate of user $k$ in (\ref{rate}) is further lower bounded by
\begin{align}  \label{rate_accurate_lowerbound}
R_{k} &\geq  \log _{2} \left(1+\frac{p\left(M-K\right) }{\sigma^{2}(\delta+1)\left[\mathbf{\Lambda}^{-1}\right]_{kk}}   \right)\\\label{rate_lb}
&\approx \log _{2} \left(1+\frac{p\left(M-K\right) }{\sigma^{2} }  	\Big(\frac{N\alpha_k\beta}{\delta+1}+   \gamma_k\Big) \right), \text{ as } N\to\infty.
\end{align}
which scales on the order of $\mathcal{O}\left(\log_{2}\left(MN\right)\right)$.
\end{corollary}

	\itshape \textbf{Proof:}  \upshape Since we consider the existence of direct links, there is $\mathbf{\Omega}_d\succ \mathbf{0}$ and then $\mathbf{\Lambda}\succ \mathbf{0}$ and $\mathbf{\Lambda}^{-1}\succ \mathbf{0}$. Besides, we have $\mathbf{\Lambda}^H=\mathbf{\Lambda}$. Based on the Woodbury's identity, we have
	\begin{align}\label{equality}
&{\left[\left(\mathbf{\Lambda}+\beta \delta \mathbf{H}_{1}^{H} \mathbf{\Phi}^{H} \mathbf{a}_{N} \mathbf{a}_{N}^{H} \mathbf{\Phi} \mathbf{H}_{1}\right)^{-1}\right]_{k k}} \qquad\qquad\qquad\qquad\qquad\qquad\qquad\qquad\qquad\nonumber\\
&\!=\!\!\left[\mathbf{\Lambda}^{-1}\right]_{k k}-\frac{\beta \delta\left[\mathbf{\Lambda}^{-1} \mathbf{H}_{1}^{H} \mathbf{\Phi}^{H} \mathbf{a}_{N} \mathbf{a}_{N}^{H} \mathbf{\Phi}\mathbf{H}_{1} \mathbf{\Lambda}^{-1}\right]_{k k}}{1+\beta \delta  \mathbf{a}_{N}^{H} \mathbf{\Phi} \mathbf{H}_{1} \mathbf{\Lambda}^{-1} \mathbf{H}_{1}^{H} \mathbf{\Phi}^{H} \mathbf{a}_{N}} \\\label{inequality}
&\!=\!\!\left[\mathbf{\Lambda}^{-1}\!\right]_{k k}\!\!-\!\frac{\beta \delta\left|  \left[\mathbf{\Lambda}^{-1} \mathbf{H}_{1}^{H} \mathbf{\Phi}^{H} \mathbf{a}_{N} \right]_{k}  \right| ^2}{1\!+\!\beta \delta  \mathbf{a}_{N}^{H} \mathbf{\Phi} \mathbf{H}_{1} \mathbf{\Lambda}^{-1} \mathbf{H}_{1}^{H} \mathbf{\Phi}^{H} \mathbf{a}_{N}} \!\leq\!\! \left[\mathbf{\Lambda}^{-1}\right]_{k k}\!.
	\end{align}
	Substituting (\ref{inequality}) into (\ref{rate}), we arrive at (\ref{rate_accurate_lowerbound}).
	Note that the $k$-th diagonal element of $\mathbf{H}_1^H\mathbf{H}_1$ equals ${\alpha_k} N$, while the non-diagonal elements are not proportional to $N$. When $N\to\infty$, we can approximate $\mathbf{H}_1^H\mathbf{H}_1$ as $N \mathrm{diag}\left\{\alpha_1,\ldots,\alpha_K\right\} $, which results in the approximation in (\ref{rate_lb}). \hfill $\blacksquare$

Corollary \ref{corollary4} reveals a very promising capacity gain. It is well-known that the ergodic rate of RIS-aided systems scales as $\mathcal{O}\left(\log_{2}\left(MN^2\right)\right)$ in the single-user scenario\cite{han2019large}. Here, we prove that by using ZF detectors in setup of multiple users, the rate of each user could still scale as $\mathcal{O}\left(\log_{2}\left(MN\right)\right)$, which demonstrates that the considered systems can achieve a promising sum user rate.
Besides, comparing (\ref{rate_lb}) with (\ref{rate_conventional}), it is shown that the RIS-aided massive MIMO systems with ZF detectors always outperform RIS-free massive MIMO systems. Meanwhile, it can be observed that the lower bound (\ref{rate_lb}) tends to (\ref{rate}) when $\delta\to0$ and tends to (\ref{rate_conventional}) when $\delta\to\infty$.

Next, we design the RIS phase shifts based on (\ref{rate}), which depends only on the statistical CSI. The sum-rate maximization problem can be formulated as follows
\begin{subequations}\label{p1}
	\begin{equation}\label{objective1}
	\max\limits_{\mathbf{\Phi}}  \;\; R^s= \sum\nolimits_{k=1}^{K} R_{k},\qquad\quad
	\end{equation}
	\begin{equation}
	\text {s.t. } \;\;  \left|[\mathbf{\Phi}]_{nn}\right| =1,\;\;1\leq n\leq N. \label{constraint1}
	\end{equation}
\end{subequations}

	Problem (\ref{p1}) is non-convex due to the non-convex unit modulus constraint. However, we can still obtain a sub-optimal solution based on the gradient ascent method. For tractability, we rewrite $\mathbf{\Phi} = {\rm diag}\left\{\mathbf{v}^H\right\}$, where $\mathbf{v}=[e^{j \theta_1},\ldots,e^{j \theta_N} ]^H$. Then, we provide the gradient vector with respect to $\mathbf{v}$ in the following lemma.
\begin{lem}
The gradient of the objective function in (\ref{p1}) is
\begin{align}
\frac{\partial R^s(\mathbf{v})}{\partial \mathbf{v}^{*}}=\sum_{k=1}^{K} \frac{\frac{\mathbf{B} \mathbf{v}}{\mathbf{v}^{H} \mathbf{A}_{k} \mathbf{v}}-\frac{\mathbf{v}^{H} \mathbf{B} \mathbf{v} \mathbf{A}_{k} \mathbf{v}}{\left(\mathbf{v}^{H} \mathbf{A}_{k} \mathbf{v}\right)^{2}}}{\ln (2)\left(1+\frac{\mathbf{v}^{H} \mathbf{B} \mathbf{v}}{\mathbf{v}^{H} \mathbf{A}_{k} \mathbf{v}}\right)},
\end{align}
where
\begin{align}
&\mathbf{A}_{k}=\frac{ \sigma^{2}(\delta+1) }{ p(M-K) }\left(\left[\mathbf{\Lambda}^{-1}\right]_{k k} \mathbf{B}-\beta \delta \mathbf{s}_{k} \mathbf{s}_{k}^{H}\right),\\\label{Bk}
&\mathbf{B}=\frac{1}{N} \mathbf{I}_{N}+\beta \delta \operatorname{diag}\left(\mathbf{a}_{N}^{H}\right) \mathbf{H}_{1} \mathbf{\Lambda}^{-1} \mathbf{H}_{1}^{H} \operatorname{diag}\left(\mathbf{a}_{N}\right),
\end{align}
with $\mathbf{s}_{k}^H \triangleq \left[\mathbf{\Lambda}^{-1} \mathbf{H}_{1}^{H} \operatorname{diag}\left(\mathbf{a}_{N}\right)\right]_{({k,:})}$ corresponds to the $k$-th row vector.

\end{lem}

\itshape \textbf{Proof:}  \upshape Substituting $\mathbf{\Phi}^{H} \mathbf{a}_{N} =\operatorname{diag}\left(\mathbf{a}_{N}\right) \mathbf{v} $ into (\ref{equality}) and utilize (\ref{Bk}), we have
\begin{align}\label{inverse}
&{\left[\left(\mathbf{\Lambda}+\beta \delta \mathbf{H}_{1}^{H} \mathbf{\Phi}^{H} \mathbf{a}_{N} \mathbf{a}_{N}^{H} \mathbf{\Phi} \mathbf{H}_{1}\right)^{-1}\right]_{k k}} \nonumber\\
&=\! \left[\!\mathbf{\Lambda}^{-1}\!\right]_{k k} \!-\! \frac{\beta \delta\!\left[\!\mathbf{\Lambda}^{-1} \mathbf{H}_{1}^{H} \operatorname{diag}\!\left(\mathbf{a}_{N}\right)\! \mathbf{v} \mathbf{v}^{H} \operatorname{diag}\!\left(\mathbf{a}_{N}^{H}\right) \mathbf{H}_{1} \mathbf{\Lambda}^{-1}\!\right]_{k k}}{1+\beta \delta \mathbf{v}^{H} \operatorname{diag}\!\left(\mathbf{a}_{N}^{H}\right) \mathbf{H}_{1} \mathbf{\Lambda}^{-1} \mathbf{H}_{1}^{H} \operatorname{diag}\left(\mathbf{a}_{N}\right) \mathbf{v}} \nonumber\\
&=\! \frac{\mathbf{v}^{H}\left\{\left[\mathbf{\Lambda}^{-1}\right]_{k k} \mathbf{B}-\beta \delta \mathbf{s}_{k} \mathbf{s}_{k}^{H}\right\} \mathbf{v}}{\mathbf{v}^{H} \mathbf{B} \mathbf{v}}.
\end{align}

Then, substituting (\ref{inverse}) into (\ref{rate}), the sum rate can be rewritten as $R^{s}=\sum_{k=1}^{K} \log _{2}\left(1+\frac{\mathbf{v}^{H} \mathbf{B} \mathbf{v}}{\mathbf{v}^{H} \mathbf{A}_{k} \mathbf{v}}\right)$. Based on the chain rule, the gradient of a real function with respect to complex vector variable is given by\cite{hunger2007introduction}
\begin{align}\label{gradient_total}
\frac{\partial R^{s}\left(\mathbf{v}\right)}{\partial \mathbf{v}^{*}}=\sum\limits_{k=1}^{K} \frac{1}{\ln (2)\left(1+\frac{\mathbf{v}^{H} \mathbf{B} \mathbf{v}}{\mathbf{v}^{H} \mathbf{A}_{k} \mathbf{v}}\right)} \frac{\partial\left(\frac{\mathbf{v}^{H} \mathbf{B} \mathbf{v}}{\mathbf{v}^{H} \mathbf{A}_{k} \mathbf{v}}\right)}{\partial \mathbf{v}^{*}}.
\end{align}

Using $\frac{\partial\left\{\mathbf{v}^{H} \mathbf{B} \mathbf{v}\right\}}{\partial \mathbf{v}^{*}}=\mathbf{B} \mathbf{v}$, and $\frac{\partial\left\{\mathbf{v}^{H} \mathbf{A}_{k} \mathbf{v}\right\}}{\partial \mathbf{v}^{*}}=\mathbf{A}_{k} \mathbf{v}$, we have
\begin{align}\label{gradient_part}
\begin{aligned}
&\frac{\partial\left(\frac{\mathbf{v}^{H} \mathbf{B} \mathbf{v}}{\mathbf{v}^{H} \mathbf{A}_{k} \mathbf{v}}\right)}{\partial \mathbf{v}^{*}}=\frac{\left\{\frac{\partial\left(\mathbf{v}^{H} \mathbf{B} \mathbf{v}\right)}{\partial \mathbf{v}^{*}}\right\} \mathbf{v}^{H} \mathbf{A}_{k} \mathbf{v}-\mathbf{v}^{H} \mathbf{B} \mathbf{v}\left\{\frac{\partial\left(\mathbf{v}^{H} \mathbf{A}_{k} \mathbf{v}\right)}{\partial \mathbf{v}^{*}}\right\}}{\left(\mathbf{v}^{H} \mathbf{A}_{k} \mathbf{v}\right)^{2}} \\
&=\frac{\mathbf{B v} \mathbf{v}^{H} \mathbf{A}_{k} \mathbf{v}-\mathbf{v}^{H} \mathbf{B v} \mathbf{A}_{k} \mathbf{v}}{\left(\mathbf{v}^{H} \mathbf{A}_{k} \mathbf{v}\right)^{2}} =\frac{\mathbf{B v}}{\mathbf{v}^{H} \mathbf{A}_{k} \mathbf{v}}-\frac{\mathbf{v}^{H} \mathbf{B} \mathbf{v} \mathbf{A}_{k} \mathbf{v}}{\left(\mathbf{v}^{H} \mathbf{A}_{k} \mathbf{v}\right)^{2}}.
\end{aligned}
\end{align}

Substituting (\ref{gradient_part}) into (\ref{gradient_total}) completes the proof.
\hfill $\blacksquare$

Assume the variable in the $t$-th iteration is $\mathbf{v}^t$.
Then,  the next variable $\mathbf{v}^{t+1}$ in the $(t+1)$-th iteration is given by
\begin{align}
&\tilde{\mathbf{v}}^{t+1}=\mathbf{v}^{t}+\mu \left.\frac{\partial R^{s}\left(\mathbf{v}\right)}{\partial \mathbf{v}^{*}} \right|_{\mathbf{v}=\mathbf{v}^t},\\\label{project}
&\mathbf{v}^{t+1}=\exp \left(j \arg \left(    \tilde{\mathbf{v}}^{t+1}   \right)\right),
\end{align}
where $\mu$ is the step size which can be chosen by using backtracking line search\cite{kammoun2020asymptotic,papa2021maxmin}. (\ref{project}) is a projection operation for meeting the unit modulus constraint (\ref{constraint1}).


\section{Simulation Results}
Unless otherwise stated, we consider $K=4$ users evenly located on the half-circle centered of an RIS with a radius $d_{UI}=20$ m. The distance between the RIS and the BS is $d_{IB}=700$ m. Using $d_{UI}$ and $d_{IB}$, the distance between the users and the BS can be calculated by their geometric relationship as \cite{zhi2020directLinks}. Based on the distances, the path loss factors $\alpha_k$, $\beta$ and $\gamma_k$ are calculated the same as \cite{zhi2020directLinks}. Besides, we set $M=N=64$, $p=30$ dBm, $\delta=1$ and $\sigma^2=-104$ dBm. The angles in the LoS channels are generated randomly from $[0,2\pi]$. The Monte Carlo simulations are obtained based on (\ref{rate_definitation}) with $10^4$ times average.

\begin{figure*}
	\centering
	\begin{minipage}[t]{0.33\linewidth}
		\centering
		\includegraphics[width=2.5in]{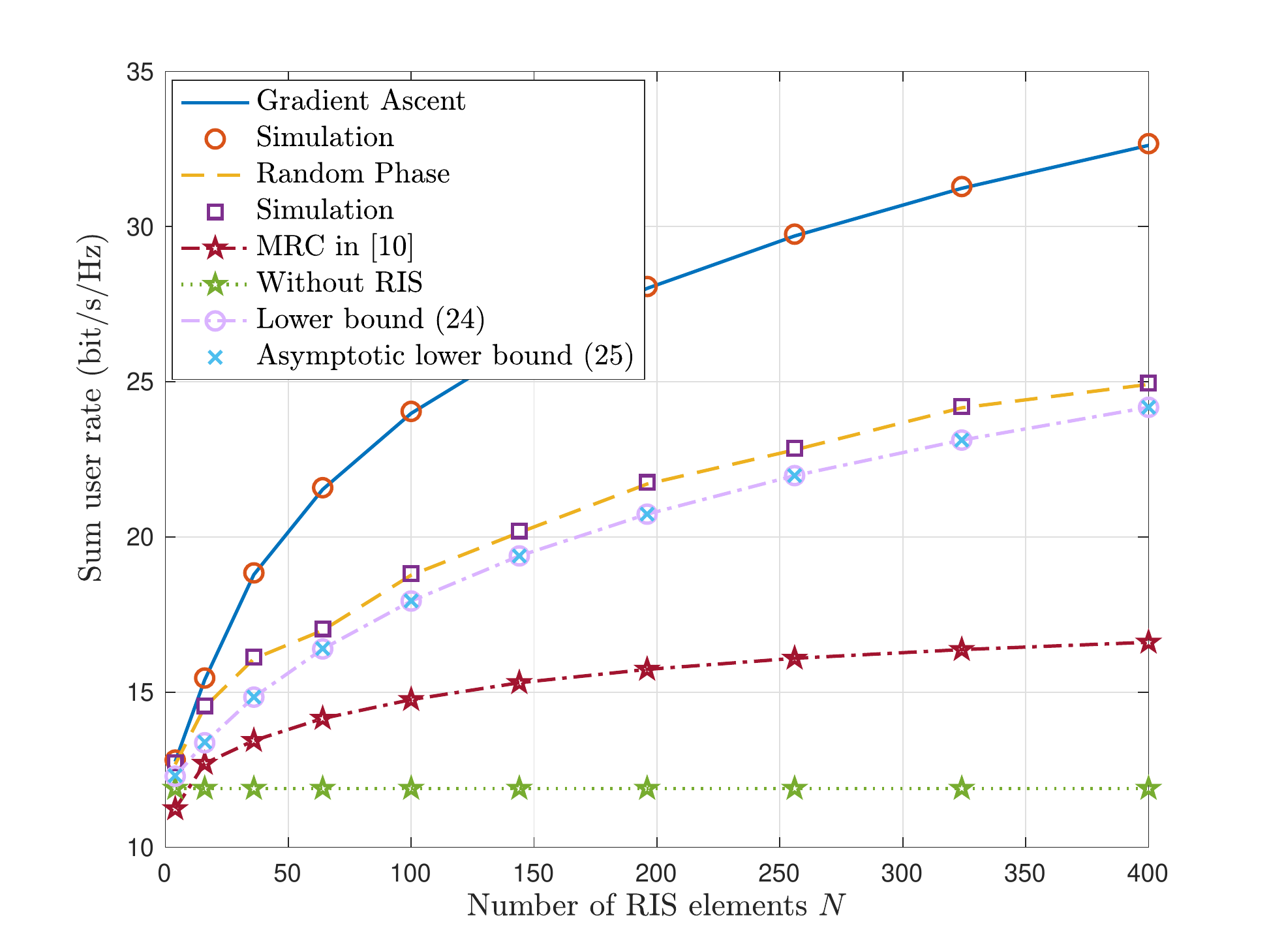}
		\caption{Rate versus the number of RIS \\elements  $N$. \qquad}
		\label{figure2}
	\end{minipage}%
	\begin{minipage}[t]{0.33\linewidth}
		\centering
		\includegraphics[width=2.5in]{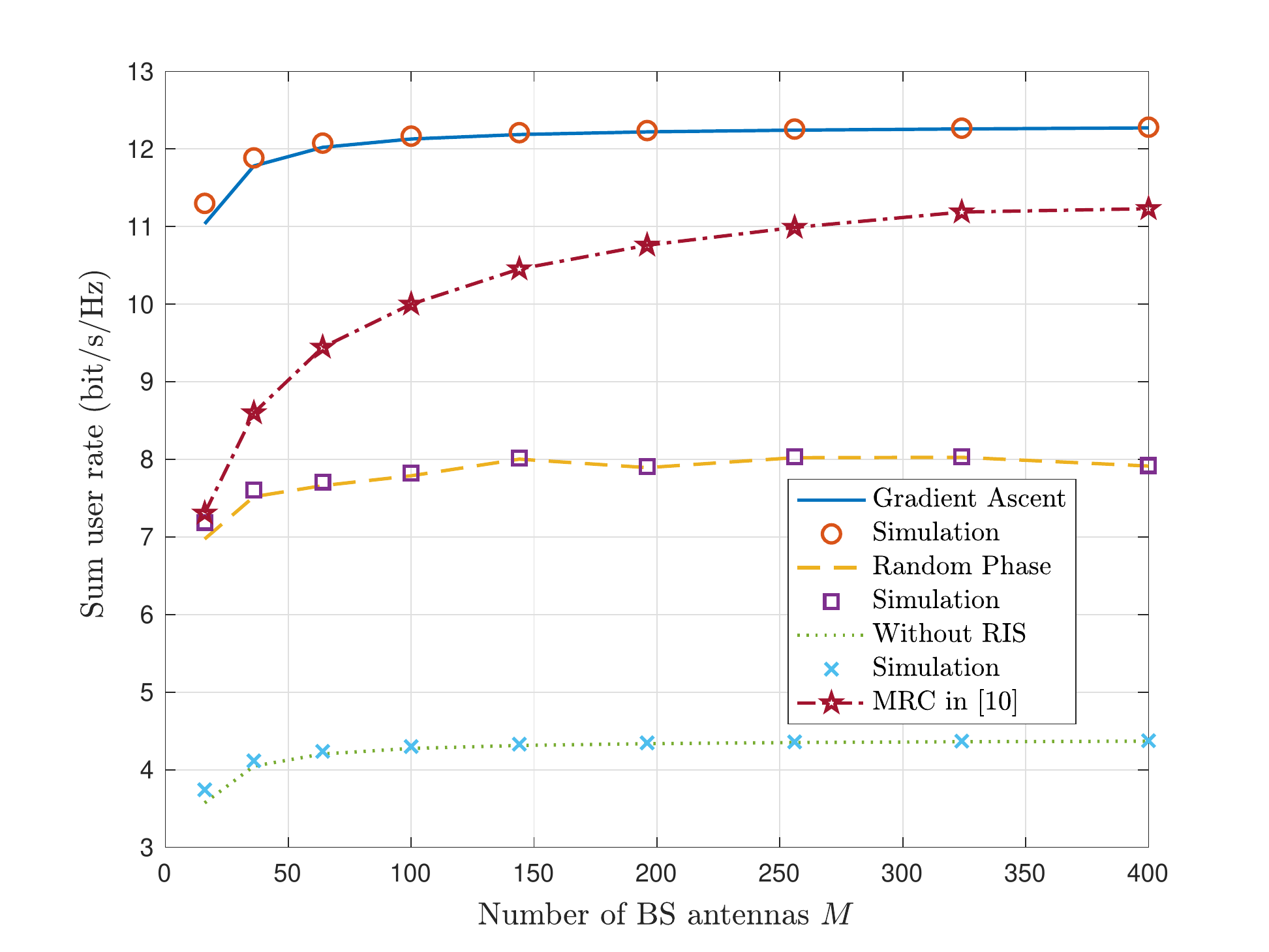}
		\caption{Rate versus $M$, where power is scaled \\down as $p=10/M$.}
		\label{figure3}
	\end{minipage}
	\begin{minipage}[t]{0.33\linewidth}
		\centering
		\includegraphics[width=2.5in]{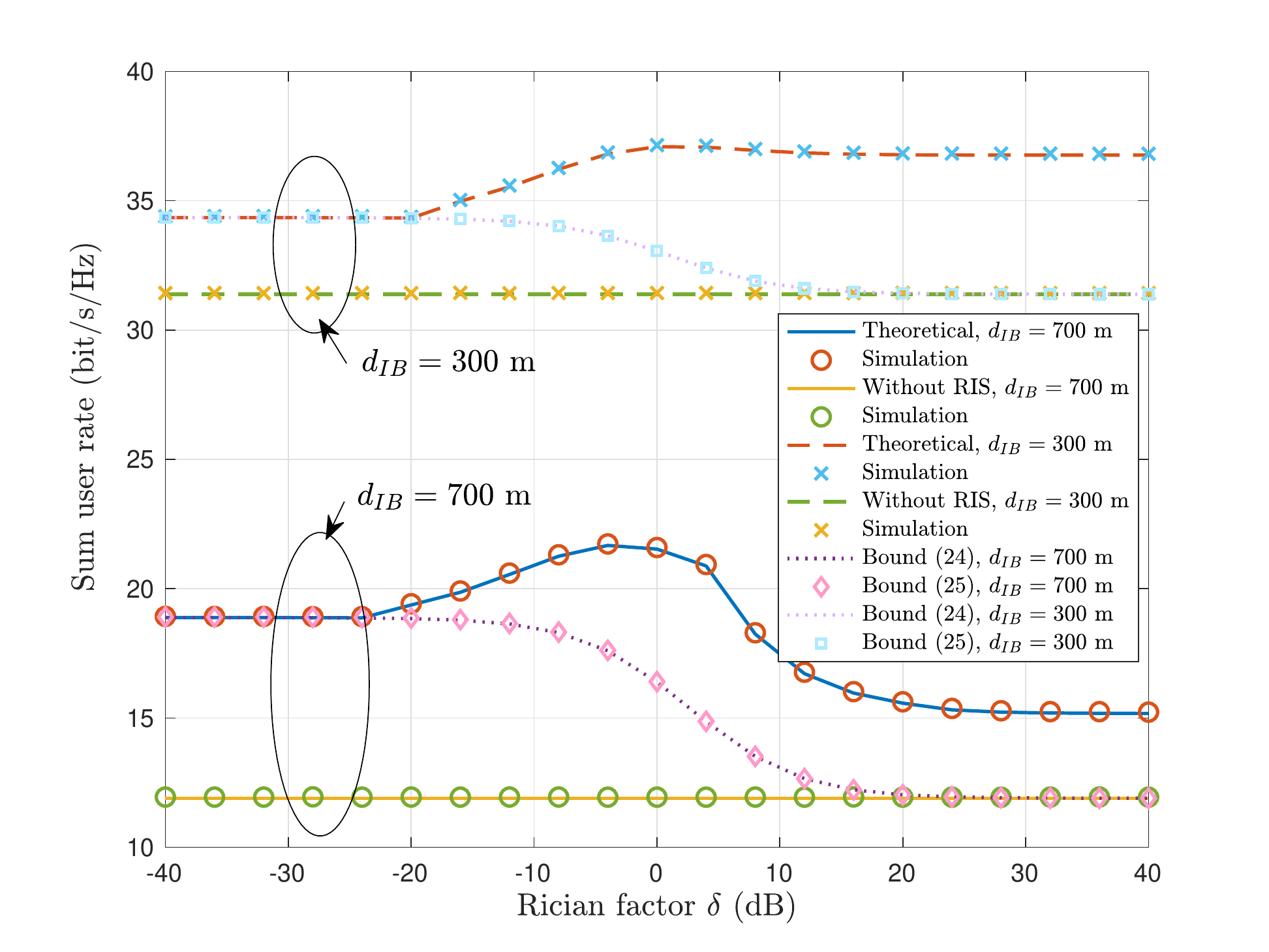}
		\caption{Rate versus the Rician factor $\delta$.}
		\label{figure4}
	\end{minipage}
\end{figure*}

In Fig. \ref{figure2}, it can be observed that  ZF-based RIS design outperforms the random phase shifts-based design, the MRC-based design, and the RIS-free systems.
	The superiority of ZF over MRC lies in the fact that RIS-aided systems suffer from severe multi-user interference\cite{zhi2020directLinks}. This is because users share the same RIS-BS channel and then their cascaded channels are highly correlated. Therefore, by effectively eliminating the interference, ZF can achieve a higher ergodic rate than MRC.
	In addition, Fig. \ref{figure2} validates the accuracy of the approximate in (\ref{rate_lb}).
In Fig. \ref{figure3}, we verify the power scaling law as expected in Corollary \ref{corollary1}. It again emphases the advantages of ZF-based RIS systems. Besides, all numerical results show that the derived lower bound (\ref{rate}) is very tight with the Monte Carlo simulations.

%
%
%

Fig. \ref{figure4} plots the rate versus Rician factor $\delta$. It can be seen that when $d_{IB}=700$ m, the rate with large $\delta$ performs worse, while a contrary result is observed when $d_{IB}=300$ m. This is because when $d_{IB}$ is large, the direct channel becomes weak, and then the cascaded channel $\mathbf{G}$ becomes a dominant factor. In this case, when $\delta$ is large, the channel $\mathbf{G}$ becomes rank-deficient, which degrades the rate. However, when $d_{IB}$ is small, the direct links are strong. Since the direct links have full-rank, the aggregate channel could always have full-rank. As shown in (\ref{rate}), the variable $\mathbf{\Phi}$ can play more roles when $\delta$ is large, which results in large performance gains.
Furthermore, Fig. \ref{figure4} validates the tightness of (\ref{rate_lb}) under all Rician factors.

\section{Conclusion}
An RIS-aided massive MIMO system with ZF detectors was considered in this paper.
 We first derived the closed-form ergodic rate expressions, whose lower bound demonstrates that the rate can scale on the order of $\mathcal{O}\left(\log_{2}\left(MN\right)\right)$. Then, using the gradient ascent algorithm, we optimized the phase shifts of the RIS based on statistical CSI. Finally, simulation results validated the correctness of our analytical results.

\bibliographystyle{IEEEtran}
\vspace{-6pt}
\bibliography{myref.bib}
\end{document}